\documentclass[12pt]{article}
\usepackage{amsmath}
\usepackage{cite}
\usepackage{epsfig}
\usepackage{epstopdf}
\usepackage{amssymb}

\def\[{\left [}
\def\]{\right ]}
\def\({\left (}
\def\){\right )}
\def\del{\partial}
\def\L{\Lambda}

\def\r{\rho}
\def\oo{{\mathcal O}}

\def\m{\mu}\def\n{\nu}

\def\e{\varepsilon}

\def\r2{\sqrt{2}}

\def\del{\partial}
\def\sech{{\rm sech}}
\def\bra{{\langle}}
\def\ket{{\rangle}}

\newcommand{\bbibitem}[1]{\bibitem{#1}\marginpar{#1}}

\def\Label#1{\label{#1}%
  \smash{\hbox to0pt{\raise1ex\hbox{\tiny[#1]}\hss}}}
\def\noLabels{\let\Label=\label}
\def\nobbibitem{\let\bbibitem=\bibitem}

\newcommand{\be}{\begin{equation}}
\newcommand{\ee}{\end{equation}}
\newcommand{\bea}{\begin{eqnarray}}
\newcommand{\eea}{\end{eqnarray}}

\newcommand{\beq} {\begin{equation}}
\newcommand{\eeq} {\end{equation}}
\newcommand{\beqa} {\begin{eqnarray}}
\newcommand{\eeqa} {\end{eqnarray}}

\newcommand{\beqn}{\begin{eqnarray}}
\newcommand{\eeqn}{\end{eqnarray}}

\def\bra{\langle}
\def\ket{\rangle}

\begin{document}

\begin{flushright}
HIP-2009-29/TH
\end{flushright}

\vskip 2cm \centerline{\Large {\bf Inhomogeneous Structures in
    Holographic Superfluids:}}
 \vskip 2mm
\centerline{\Large {\bf I. Dark Solitons}}
 \vskip 1cm
\renewcommand{\thefootnote}{\fnsymbol{footnote}}
\centerline
{{\bf Ville Ker\"anen,$^{1}$
\footnote{ville.keranen@helsinki.fi}
Esko Keski-Vakkuri,$^{1}$\footnote{esko.keski-vakkuri@helsinki.fi}
Sean Nowling,$^{1,2}$\footnote{sean.nowling@helsinki.fi}
K. P. Yogendran,$^{1}$ \footnote{yogendran.kalpat@helsinki.fi}
}}
\vskip .5cm
\centerline{\it
${}^{1}$Helsinki Institute of Physics }
\centerline{\it P.O.Box 64, FIN-00014 University of
Helsinki, Finland}\centerline{\it
${}^{2}$Department of Mathematics and Statistics}
\centerline{\it P.O.Box 68, FIN-00014 University of
Helsinki, Finland}

\setcounter{footnote}{0}
\renewcommand{\thefootnote}{\arabic{footnote}}

\begin{abstract}
We begin an investigation of inhomogeneous structures in holographic
superfluids. As a first example, we study domain wall like defects in
the 3+1 dimensional Einstein-Maxwell-Higgs theory, which was developed
as a dual model for a holographic superconductor.  In
\cite{Keranen:2009vi}, we reported on such ``dark solitons'' in
holographic superfluids.  In this work, we present an extensive
numerical study of their properties, working in the probe limit. We
construct dark solitons for two possible condensing operators, and
find that both of them share common features with their standard
superfluid counterparts. However, both are characterized by two
distinct coherence length scales (one for order parameter, one for
charge condensate). We study the relative charge depletion factor and
find that solitons in the two different condensates have very distinct
depletion characteristics. We also study quasiparticle excitations
above the holographic superfluid, and find that the scale of the
excitations is comparable to the soliton coherence length scales.
\end{abstract}

\newpage


\section{Introduction}
 
The AdS/CFT correspondence
\cite{Maldacena:1997re,Gubser:1998bc,Witten:1998qj} -- or
gauge-gravity duality, as its more phenomenological incarnation is
called -- is being used as a tool to study an increasing variety of
strongly interacting systems. When the system is also "close" to being
conformal, it is natural to try to construct a holographic gravity
dual model.  The main practical advantage of holography is that it
allows a new, computationally simpler framework to explore quantum
systems which are outside the reach of conventional perturbative
techniques. The situation of interest in this document will be a
system in which a U(1) symmetry is spontaneously broken, as is
relevant for superfluidity and superconductivity.

Superfluidity may occur in a system at low temperatures when the
ground state becomes occupied by a macroscopic number of particles.
This is known to occur in both interacting bosonic systems
(Bose-Einstein condensates, BEC) and interacting fermions with
Bardeen-Cooper-Schrieffer pairing (BCS).  In fact, cold atomic systems
can display both BEC-like and BCS-like superfluidity and even a smooth
transition between them.

It is not at all obvious what kinds of superfluids may be realized
holographically. 
A prototype example is the recently constructed
gravitational dual theory \cite{Hartnoll:2008vx}, building on the
model \cite{Gubser:2008px}, for the purpose or modeling BCS
superconductors \cite{Hartnoll:2008vx,Hartnoll:2008kx} or relativistic
superfluids with a spontaneously broken global U(1) symmetry
\cite{Herzog:2008he}. The model was also extended to an effective
theory for some properties of a class of quantum Hall fluids
\cite{KeskiVakkuri:2008eb}.
Typically, in systems that are dual to AdS models, one finds both
fermionic and bosonic excitations.  Thus, one might hope to be able to
holographically model both BEC-like and BCS-like superfluidity, as
well as more complicated fluids with mixed behavior. For closer
contact with real world cold atom systems, a required ingredient would
also be manifest non-relativistic symmetry, for which the first
holographic models were studied in
\cite{Son:2008ye,Balasubramanian:2008dm}. In this paper, and in the
associated sequence of works to follow, we will focus on studying
extended configurations in the model \cite{Hartnoll:2008vx}, and we
will find a richer variety of possibilities than previously expected.

Hydrodynamic properties of the relativistic holographic superfluid have been
explored in \cite{Amado:2009ts,Herzog:2009md} -- the latter in
particular argues that the superfluid does not obey the Landau
relationship between sound velocities. In \cite{Maeda:2009wv}, masses
of quasiparticles were determined analytically near $T_c$ and
substantially extended in \cite{Umeh:2009ea} (to other values of the
mass of the bulk scalar field as well).

Another basic fact is that conventional superfluids are known to
support long lived spatially inhomogeneous configurations with
nontrivial topology.
An experimentally and theoretically interesting class of defects,
found in superfluids, are domain wall like defects called dark
solitons. A dark soliton is an interface of reduced (charge) density
between two superfluid phases, with the order parameter changing sign
across the interface. Holographic counterparts were recently reported
by us in \cite{Keranen:2009vi}.  Apart from curiosity, there are
several reasons to study such solitons, a primary one being to explore
their properties at strong coupling, a difficult problem using
traditional methods.  Although the gravity dual has different
symmetries and microscopic properties from the real life cold atomic
systems, there may be some universal features.  At the very least it
provides a ``spherical cow'' example allowing one to study properties
of a class of dark solitons at strong coupling.

The holographic dark solitons are also interesting solutions in their
own right -- they turn out to reveal important information about the
different ways of condensing operators in \cite{Hartnoll:2008vx} and
about the nature of the holographic superfluid. First of all, we find
that the solitons have two distinct length scales: one for
the charge density and one for the order parameter. This behavior is
in contrast to a single scale found in simple solutions to the
Gross-Pitaevskii equation for superfluids. Furthermore, by studying
the degree of charge density depletion we discover that the two
different choices for the condensing operator, discussed in
\cite{Hartnoll:2008vx}, lead to different properties for the
associated superfluids.
We will also compare the length scales as set by these solitonic
configurations with the microscopic masses of quasiparticles.

This document is organized as follows. We begin with a brief
discussion of the holographic description of a superfluid.  The
configurations of interest are obtained by solving a system of partial
differential equations in AdS space. These equations and the boundary
conditions -- which sustain the soliton -- are presented in Section
\ref{HoloEqns}.  However, the equations seem intractable
analytically. We therefore have to numerically solve the
equations. The discretization of the equations and the method we use
to solve the equations are presented in Section \ref{Numerics}. In
order to identify properties of interest to be obtained from the
numerical solutions, we will briefly discuss the Gross-Pitaevskii
equations in Section \ref{Dark Solitons}. We will then present our results for
the {\em holographic} dark solitons and conclude with a discussion of
the results and future directions.

\section{Holographic Description \label{HoloEqns}}
An explicit holographic modeling of a superfluid system was
constructed in \cite{Hartnoll:2008vx} following the ideas in
\cite{Gubser:2008px} (closely related ideas also appeared in
\cite{Basu:2008st}). One considers a system consisting of a complex
scalar field $\Psi$ interacting with a U(1) gauge field
$F_{\mu\nu}=\del_\mu A_\nu-\del_\nu A_\mu$ in 3+1 dimensions in the
presence of gravity with a cosmological constant $\L$. The action for
this system is
\be
{\mathcal S}=\int d^3x dz
\sqrt{-g}\[\frac{1}{2\kappa^2_4}(R-12\Lambda)+\frac{1}{q^2}
\left(-\frac{1}{4} F_{\mu\nu}
F^{\mu\nu}-m^2\Psi\bar\Psi-D_\mu\Psi D^\mu \bar\Psi\right)
\]\label{action}
\ee
where the covariant derivative $D_\mu\Psi=(\del_\mu-i A_\mu)\Psi$ and
$q$ may be identified with the charge of the scalar field\footnote{We
  have rescaled the fields such that $q$ only appears in front of the
  matter action.}. Focusing on just these degrees of freedom (rather
than keeping all the fields required for a UV complete gravity
theory), amounts to concentrating on the most relevant operators of
the field theory. One could include other fields (say, a fermion field
which would be dual to fermionic quasiparticles). In keeping with the
usual terminology, we will refer to the fields of the gravitational
theory as the ``bulk'' fields in what follows.

The AdS/CFT correspondence \cite{Maldacena:1997re} requires us to
find solutions to the equations of motion of the 4D-gravitating
system, such that the metric is asymptotically AdS. The properties of
the dual field theory are then ``read off'' from the asymptotic
behavior of the various fields of this gravitating system
(\cite{Gubser:1998bc,Witten:1998qj}, see the review
\cite{Aharony:1999ti}).

In our case, since we are trying to study a spatially inhomogeneous
configuration, solving the full gravitating system is a difficult
problem. Therefore, we consider a limit in which the energy density in
the complex scalar field and the Maxwell field (in 4D) is scaled to
zero. This may be achieved by taking $\frac{\kappa_4}{q}\rightarrow 0$
keeping the field values $\Psi, A_\mu$ finite. In this limit (termed
the probe limit in the AdS literature) the metric is determined,
independent of the matter fields, to be a 4-D planar AdS black hole
with a metric
\be
ds^2 = L^2(-\frac{f dt^2}{z^2}+\frac{dz^2}{f z^2}+\frac{
  d\vec{x}^2}{z^2}),
\ \ \ f(z)=1-\left(\frac{z}{z_T}\right)^3.
\ee
The Hawking temperature of the black hole, $T_H=\frac{3}{4\pi z_T}$,
is identified as the equilibrium temperature of the dual field theory.

The complex scalar field and gauge field then propagate on this
background and their dynamics is determined by an action
\be
{\mathcal S}=\int d^3x dz \sqrt{-g}\left(-\frac{1}{4} F_{\mu\nu}
F^{\mu\nu}-m^2\Psi\bar\Psi-D_\mu\Psi D^\mu \bar\Psi\right)
\ee
where the covariant derivative $D_\mu\Psi=(\del_\mu-iA_\mu)\Psi$ and
the various indices are contracted using the metric given above.

If we rescale
\be
(z,x)\to z_T(z,x) \quad A \to \frac{A}{z_T}\quad \Psi\to \frac{\Psi}{z_T},
\ee
then the new co-ordinates $z,x$ are dimensionless as are the new
fields $\Psi,A$.  Further, all dependence on $z_T$ is also removed.

The equations of motion for this system are
\beqn\label{eq:psieq}
0 &=& \frac{1}{\sqrt{-g}}D_\mu(\sqrt{-g}D^\mu \Psi)-m^2 \Psi \\
0 &=& \frac{1}{\sqrt{-g}}\partial_\mu\left(\sqrt{-g} F^{\mu\nu}\right)
+ i(\Psi\partial^\nu\bar{\Psi}-\bar{\Psi}\partial^\nu\Psi)+2A^\nu\Psi\bar{\Psi}.
\eeqn
In this work we will primarily be interested in studying static kink
solutions in the field theory and therefore the fields only have
dependence on $z$ and one spatial variable, $x$.  In the gauge where
$A_z=0$, the $A_z$ equation is satisfied if $\Psi$ is taken to be real
and $A_x=0$. We will assume translation invariance in the
$y$-direction and set $A_y=0$. Then the only nonzero component of the
gauge field is $A_0=A(z,x)$

It is important to note that, after gauge fixing, there is a residual
$\mathbb{Z}_2 $ symmetry under which the scalar field changes
sign. When studying dark soliton solutions to the equations of motion,
the scalar field changes sign as a function of $x$,
$\lim_{x\rightarrow \pm \infty} \Psi(x) = \pm |\Psi(\infty)|$.  One
might think that this sign flip may also be gauged away, but this
would require a gauge transformation that is singular as the scalar
field crosses through zero, and hence is not an allowed gauge
transformation.

It is convenient to redefine the field slightly, $\Psi = z
\tilde{R}/\sqrt{2} $, and rewrite equations of motion
\beqn\label{eq:bulk}
f\tilde R''+f' \tilde R'-z \tilde R +\partial_x ^2 \tilde R
+\tilde R (\frac{A^2}{f})=0\\
f A''+\partial_x ^2 A-\tilde R ^2 A=0
\eeqn

\subsection{Bulk-boundary dictionary}

Typically, in AdS spacetimes, the solutions to the equations of motion
may be segregated into "normalizable" and "non-normalizable" parts
determined by their leading $z$ behavior as one approaches the
boundary.  According to the AdS/CFT dictionary \cite{Gubser:1998bc,Witten:1998qj,Aharony:1999ti} 
the two have different
interpretations in the dual field theory.  The boundary values of the
non-normalizable modes are interpreted as sources for the operators of
the dual field theory.  The normalizable modes are then identified
with the vacuum expectation values of the corresponding operators
sourced by the non-normalizable modes \cite{Balasubramanian:1998de}.  
The scaling behavior of the
dual operator is also fixed and may be extracted from the asymptotic
behavior of the normalizable modes as functions of $z$.  This is
explicitly realized by identifying the generating functional of
connected correlation functions, $W[J]$, of the strongly coupled field
theory with the on-shell action for the bulk gravity theory (the bulk
fields have boundary values $\phi_\partial = J$.
(\cite{Gubser:1998bc,Witten:1998qj})

\beq
S_{\mathrm{Grav.},\ \mathrm{O.S.}}{\big |}_{\phi_\partial=J} = W_{QFT}[J].
\eeq
One simple way to see the relationship between normalizable modes and
expectation values is to consider the change in the action of the
gravity theory caused by infinitesimal changes in the
"non-normalizable" gravity modes (field theory sources).  Upon partial
integration and the inclusion of a suitable counterterm, the variation
of the action (\ref{action}) in terms of the fields $\Psi,A$, becomes
\bea \delta S&=&
-\int d^4x \del_\m(\delta{\Psi}\sqrt{g}\partial^\m \Psi)
+\frac{1}{2} \del_\m(\sqrt{g}\delta A\ F^{\m 0})\\
&&+\int d^4x \delta \Psi\  \mathcal{E}( \Psi)
+\frac{1}{2}\delta A_\n\ \mathcal{E}(
A)\nonumber\\
&&+ {\rm\ terms\ higher\ order\ in\ the\ variations}.\nonumber
\eea
We have grouped the equations of motions in (\ref{eq:psieq}) as
$\mathcal{E}(\Psi)$ and $\mathcal{E}(A)$ for brevity.  Rewriting
the surface terms as in the AdS prescription gives us
\bea\label{eq:variation}
\delta S&=&
-\int d^3x [(\delta \Psi\sqrt{g}\partial^z\Psi)+
(\sqrt{g}\delta A\ F^{z 0})]{\Big |}_{z=0}\\
&&+\int d^4x \delta\Psi\ \mathcal{E}(\Psi)
+\frac{1}{2}\delta A\ \mathcal{E}(A)
\nonumber \\&&+\
{\rm terms\ higher\ order\ in\ the\ variations}.\nonumber
\eea
contributions only from the first line in the above.

As in \cite{Hartnoll:2008vx}, we shall take the scalar field to
satisfy $m^2=-2/L^2$. For this mass value, there are two quantizations
for scalar fields in any asymptotically $\mathrm{AdS}_{4}$ spacetime
\cite{Klebanov:1999tb}. The two quantizations correspond to exchanging
the role of source and expectation value in the dual field theory.

More explicitly, it can be easily seen that, close to the boundary
at $z=0$, solutions to the equations of motion must behave as
\be
\tilde{R}\sim   \tilde R^{(1)}+z \tilde R^{(2)}+...,
\quad A \sim A ^{(0)}+z A ^{(1)}+...
\ee
in an expansion of the Frobenius type along the z-direction.  We can
regard $\tilde{R}^{(1)}$ as the source for a charged, dimension $2$
operator
\be
\tilde R^{(1)}= z_T J^{(2)} \quad
\tilde R^{(2)}= z_T ^2 \langle \oo_2\rangle\quad
A_0 ^{(0)}= z_T \mu\quad
A_0 ^{(1)}=z_T ^2\rho.
\ee
where $\langle O_2\rangle $ is the charged operator, $\mu$ is the
chemical potential, and $\rho$ is the charge density (all in the dual
field theory).  One could motivate the identification of the
asymptotic value $A_0 ^{(0)}= z_T \mu$ by noting that, in Euclidean
space, a vev of $A_0$ is naturally interpreted as a chemical potential
(since it minimally couples to a conserved charge).

In the second quantization scheme, we identify $\tilde{R}^{(2)}$ as
the source for a charged, dimension $1$ operator
\be
\tilde R^{(1)}= z_T \langle \oo_1\rangle \quad
\tilde R^{(2)}= z_T ^2 J^{(1)}\quad
A_\m ^{(0)}= z_T \mu\quad
A_\m ^{(1)}=z_T ^2\rho.
\ee
(we will use $\psi$ to denote the condensate in what follows).
The various prefactors come from the rescaling of $z$ by $z_T$ so that
the fields in the bulk are dimensionless. Solutions to the
gravitational equations of motion with nonzero non-normalizable
components are interpreted in the dual field theory as deforming the
Hamiltonian, $\delta \mathcal{H} \sim \int d^3x
J^{(i)}\mathcal{O}^{(i)}$ and would hence change the ground state
of the theory.

To obtain a unique solution to the bulk equations of motion, we need
to impose boundary and regularity conditions.  As in
\cite{Hartnoll:2008vx}, we want to study the spontaneous breaking of
the $U(1)$ symmetry unperturbed by any sources, so we will impose
boundary conditions such that in each quantization the source terms
are turned off.

In addition, we impose the following regularity conditions at the
horizon (we assume that $R$ is regular at the horizon)
\be
f' \tilde R'-z \tilde R +\partial_x ^2 \tilde R=0
\qquad A_0 (z=1,x)=0  \label{RegEqns}
\ee
(the condition on the gauge field has been argued to be a regularity
condition \cite{Gubser:2008px}).  Because the differential equations
are elliptic (\ref{eq:bulk}) outside the horizon, Dirichlet/Neumann
boundary conditions along the boundary (at $z=0$) and regularity (at
the horizon) are sufficient to fix a unique solution (a useful analogy
is with Poisson equations).

\subsection{Review of earlier results \label{recap}}

One of the features of the AdS/CFT correspondence is that global
symmetries of the field theory system appear gauged in the
gravitational theory. Given this, the $U(1)$ gauge symmetry in the
gravitational Lagrangian (\ref{action}) is interpreted as corresponding to a
global $U(1)$ symmetry of the dual field theory system. And hence, the
bulk charged scalar field $\Psi$ maps into a field theory operator
$\oo_i$ that is ``charged'' under the global $U(1)$.  We can now study
this system to see if it has ground states in which these charged
operators pick up nonzero vacuum expectation values (vevs) in the
limit where all external sources for charged operators are turned off.

In the works of \cite{Hartnoll:2008vx}, it was shown that for $\mu\geq
\mu_c$ (or equivalently small $T$, since there is only one independent
parameter $\tilde \mu=\frac{3\mu}{4\pi T}$), one can find nontrivial
solutions for the scalar field equations displaying spontaneous
symmetry breaking in the dual field theory.  Thus, for low enough
temperature (or large enough $\mu$) we obtain superfluidity (strictly,
we have only argued for a condensate, that this is indeed a superfluid
has also been established by showing the existence of a hydrodynamic
mode\cite{Herzog:2008he}). The critical value of $\mu$ (or $T$)
depends on whether we consider $\tilde{R}^{(1)}$ or $\tilde{R}^{(2)}$
as our normalizable mode (or equivalently, as the vev of the order
parameter).

\begin{figure}[height=2cm]
\begin{center}
\includegraphics{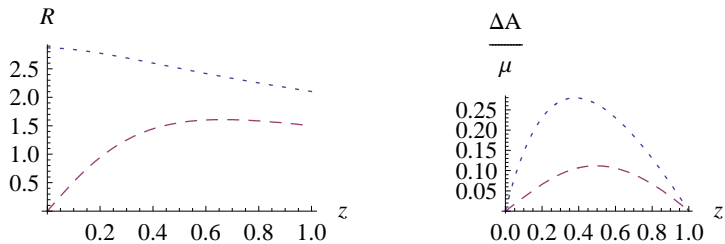}
\caption{Scalar field and Gauge field profile: The dotted line refers
  to $\oo_1$ and the dashed line to $\oo_2$ type condensate}
\label{fig:solution1}
\end{center}
\end{figure}

The graph of typical solutions look as in Fig. \ref{fig:solution1}
wherein we plot the two solutions obtained by condensing either
operator $\oo_1$ (dotted line) or $\oo_2$(dashed line)
We also note that at any temperature (or chemical potential) we have a
solution where the condensate is absent $A(z)=\tilde \mu(1-z)$ and
$\Psi=0$.  Such a solution is energetically disfavored at low
temperatures ($T<T_c$ or $\mu >\mu_c$).\cite{Herzog:2008he}

The features in the bulk gauge field are highlighted better if we
graph a subtracted potential $\Delta A(z)=A(z)/A(0)-(1-z)$ which may
be interpreted as charge density over and above the zero condensate
value.  It is seen in Fig. \ref{fig:solution1} that the bulk gauge
field profile is quite similar in the two cases.  One feature of
potential interest is that the presence of the condensate changes the
charge density by little (vide the range of the vertical axis), at
least for small chemical potentials.

We also plot the charge density and condensate as a function of the
holographic direction $z$.  One of the interesting differences between
the two types of solutions is that when $\oo_2$ condenses, the
condensate always vanishes at a point $z=z_c$.  Beyond this $z_c$
value the condensates is always monotonically decreasing as one
approaches the horizon.  For $\oo_1$ condensates, the condensate is
monotonic as a function of $z$.
\begin{figure}[ht]
\begin{center}
\includegraphics[totalheight=4cm]{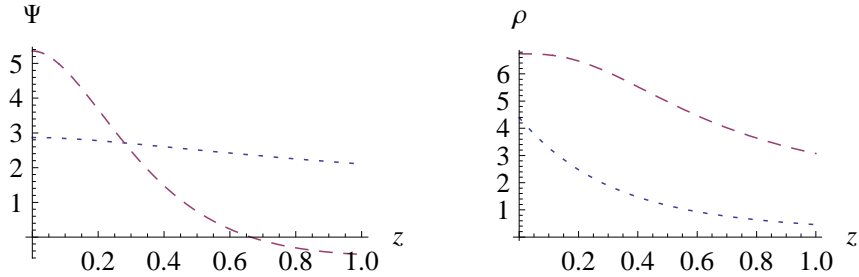}
\caption{The condensate and charge density profile: The dotted line
  refers to $\oo_1$ and the dashed line to $\oo_2$ type condensate}
\label{fig:solution3}
\end{center}
\end{figure}
The Landau-Ginzburg approach suggests that the square of the
condensate should be function of $(\mu-\mu_c)$ (\ref{tanh}).
\begin{figure}[ht]
\begin{center}
\includegraphics[totalheight=5cm]{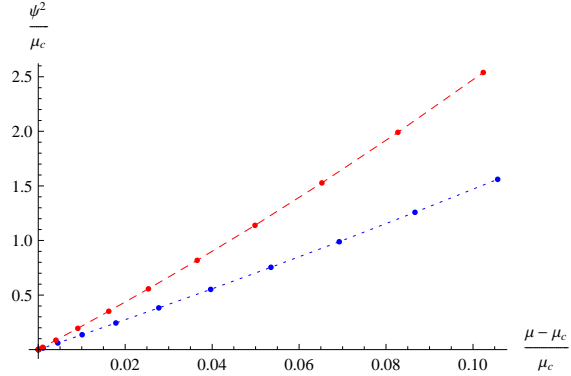}
\caption{The condensate as a function of the chemical potential: Upper
  curve is $\oo_2$, lower $\oo_1$.}
\label{fig:condmu}
\end{center}
\end{figure}
As can be seen from Fig. \ref{fig:condmu}, for small values of
$\frac{(\mu-\mu_c)}{\mu_c}$, the curve may be approximated by a
straight line.  However, for somewhat larger values, the slope changes
- and the fitting functions require higher powers.

\begin{figure}[ht]
\begin{center}
\includegraphics[totalheight=4cm]{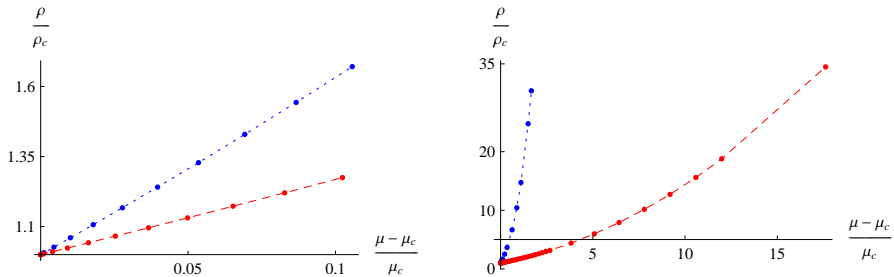}
\caption{The charge density vs. chemical potential: Upper
  curve is $\oo_2$, lower $\oo_1$.}
\label{fig:eos}
\end{center}
\end{figure}

The equation of state for this system - namely the graph of the charge
density as a function of the chemical potential for fixed temperature
looks as in the Fig. \ref{fig:eos}.
Close to the phase transition, the charge density and chemical
potential are linearly related, but near $T=0$ (large $\mu$), this is
no longer true.

One could expect that the condensate should be a linear function of
the charge density - if one imagines that the microscopic system
consists of fermions, then both the charge density and the condensate
are proportional to the vev of $\psi\psi$ - where $\psi$ is
the fermion many body wavefunction. The condensate is indeed a linear
function of the charge density at large values of the chemical
potential (i.e., close to $T=0$) but near the phase transition it
seems that increasing the condensate does not increase the charge
density by a considerable amount.
\begin{figure}[ht]
\begin{center}
\includegraphics[totalheight=4cm]{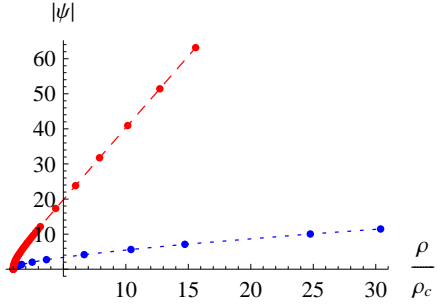}
\caption{The condensate as a function of the charge density (upper
  curve is $\oo_1$ condensate)}
\label{fig:condQ}
\end{center}
\end{figure}

\section{Numerical Methods for Dark solitons\label{Numerics}}

The bulk equations of motion (\ref{eq:bulk}) seem to be intractable
even in the homogeneous case. Therefore, we resort to numerical
methods to solve them. Since we are interested in inhomogeneous
solutions, shooting methods become unwieldy (in this regard, see
\cite{Albash:2009iq} for a tour de force of numerical work in a
closely related context). Therefore, because the differential
equations are elliptic outside the horizon, it is natural to use
relaxation methods to find solutions.

More precisely, to simulate the differential equations, we place the
system in a large box of size $1\times L_x$ in the $z$ and $x$
directions respectively.  We discretize this box using a lattice, and
then solve a discretized version of the differential equation on the
lattice using a Gauss-Seidel routine.  The Gauss-Seidel strategy is to
start with a seed configuration which obeys appropriate boundary
conditions. We can then use a discrete representation of the
differential equations to ``relax'' this configuration towards a
solution.  If the lattice is sufficiently fine, the elliptic nature of
the problem implies that the seed configuration will flow to an exact
solution to the differential equation eventually.

We set our conventions by labeling the $z$ and $x$ positions with
$i\in[0,M]$ and $j\in[0,N]$ respectively and use step sizes $h_z$ and
$h_x$.  We will denote the lattice fields as
$({\tilde R},A)\rightarrow (\tilde{R}_{i,j},A_{i,j})$.

The discretization of these equations (\ref{eq:bulk}) has to be done
separately for the interior of the lattice and the edges to at least
second order in the lattice spacings.  In the interior, when $i\neq
1,M$ $j\neq 1,N$, we can use the center difference formulae for the
derivatives,
\beqn
\frac{ \partial f(x) }{ \partial x } \Big{|}_{x=x_i} &=&
\frac{ f(x_{i+1})-f(x_{i-1})}{2h_x}+\mathcal{O}(h_x^2)
\nonumber \\
\frac{ \partial^2 f(x) }{ \partial x^2 } \Big{|}_{x=x_i} &=&
\frac{ f(x_{i+1})-2f(x_i)+f(x_{i-1}) }{ h_x^2 }+\mathcal{O}(h_x^2)
\eeqn
and therefore we get the following algebraic equations
\beqn
\tilde R_{00}&=&
\frac{((\frac{f}{h_z ^2}+\frac{f'}{2h_z })\tilde R_{+0}+(\frac{f}{h_z ^2}
-\frac{f'}{2h_z })\tilde R_{-0}+\frac{\tilde R_{0+}+\tilde R_{0-}}{h_x ^2})}
{(z-\frac{A_{00}^2}{f}+\frac{2}{h_x ^2}+\frac{2f}{h_z ^2})} \\
A_{00}&=&
\frac{\frac{f}{h_z ^2}(A_{+0}+A_{-0})+\frac{1}{h_x ^2}(A_{0+}+A_{0-})
}{
(\tilde R_{00} ^2+\frac{2}{h_x ^2}+\frac{2f}{h_z ^2})
}\label{GSeqns}
\eeqn
The above equations are to be understood as follows. At a site $i,j$,
the value of the scalar field $R_{i,j}$ which we label as $R_{00}$
is determined in terms of its neighbors $R_{i\pm 1,j}=R_{\pm 0}$ etc
by the above equations. In addition, $z$ refers to the lattice
value $z_i$, and $f$ refers to $f(z_i)$ and $f'$ must be replaced
by the center difference formula at the lattice site.

At the spatial edges (in the bulk) when $x=\pm L_x/2$ and $z\neq 0,1$,
we simply use one-sided representations of the finite difference
derivatives and again impose the equations of motion.  We have also
checked that one could also employ Neumann boundary conditions at
$x=\pm \frac{L}{2}$ without changing the numerically determined
solutions, as long as $L$ is sufficiently large.

At the horizon, we impose regularity conditions (\ref{RegEqns})
appropriately discretized. At the boundary of AdS space, we have to
numerically impose the boundary conditions
\begin{eqnarray}
&&\oo_1  \, {\rm case}: A(z=0,x)=\mu={\rm constant},\quad
\frac{\del \tilde R}{\del z}(z=0,x)=0\\\nonumber
&&\oo_2 \, {\rm case}: A(z=0,x)=\mu={\rm constant},\quad \tilde R(z=0,x)=0.
\end{eqnarray}


By cycling through the lattice, imposing either the equations of
motion, regularity, or boundary conditions depending upon the
position, an initial seed configuration relaxes to a solution to the
algebraic equations satisfying the appropriate boundary conditions.

As a first check of our algorithm, we checked that we can obtain
the symmetry breaking solutions obtained by \cite{Hartnoll:2008vx}
that correspond to having a homogeneous phase without any dark
solitons.  For the $\oo_2$ condensate for instance, we set the chemical
potential at the boundary to a fixed value, and $\tilde{R}(0,x)=0$,
and allow fairly arbitrary seed field values in the bulk of the
lattice (as well as at the horizon).  Upon subsequent iteration, we
obtain a translationally invariant solutions that match the numerical
solutions obtained by solving the corresponding one dimensional
problem using Mathematica's NDSolve to remarkable accuracy even on
fairly modest lattices.  For the $\oo_1$ condensate we enforced
$\partial_z \tilde{R}(0,x) = 0$ instead. Again, after iterating one
finds spatially homogeneous solutions which agree with
\cite{Hartnoll:2008vx}.

\subsection{Constructing Dark Solitons}

The dark solitons were constructed, numerically, as follows. We first
chose a seed configuration that asymptotes to solutions in
\cite{Hartnoll:2008vx} far away from the interface. The initial
configuration is assumed to be odd in the x-direction. We then
numerically iterate the seed configurations until it relaxes to a
stable configuration (we do this on several lattice sizes for each
soliton). In actual calculations, we are forced to use a cutoff at a
distance $\epsilon=10^{-10}$ from the horizon at $z=1$. We have
checked that this value of $\e$ does not affect the results quoted.

\begin{figure}[ht]
\begin{center}
\includegraphics[totalheight=5cm]{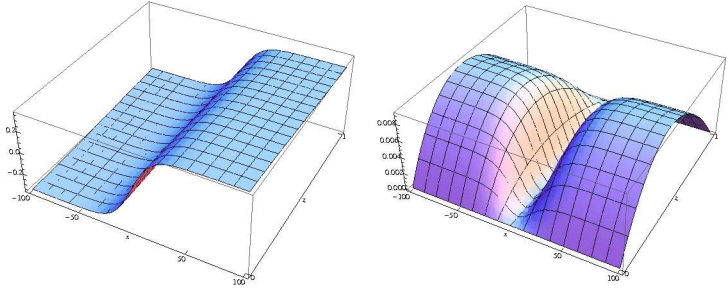}
\end{center}
\caption{$\oo_1$ case: The bulk solution (Scalar field on the left and
  subtracted gauge field on the right). }\label{fig:condo1}
\end{figure}

\begin{figure}[ht]
\begin{center}
\includegraphics[totalheight=5cm]{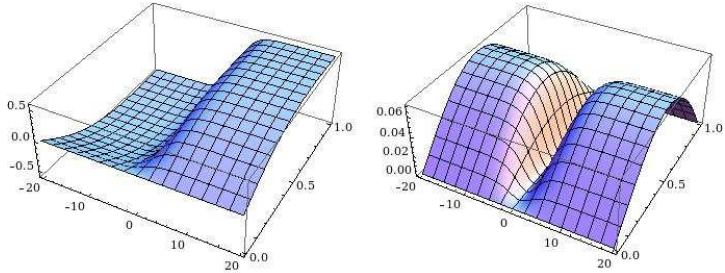}
\end{center}
\caption{$\oo_2$ case: The bulk solution (Scalar field on the left and
  subtracted gauge field on the right). }\label{fig:condo2}
\end{figure}

A typical solution so obtained is shown in Fig. \ref{fig:condo1} for the
$\oo_1$ case.
In the gauge field plot, we have subtracted a linear part
$\mu(1-z)$. Since this is an $\oo_1$ condensate, the boundary values of
the scalar field are non-vanishing.  A corresponding typical solution
for the $\oo_2$ case is shown in Fig. \ref{fig:condo2}. In this case
the boundary value of the scalar field is zero and the derivative of
the scalar field, $\partial_Z \tilde{R}$, is identified with the
condensate. 
magnitude of

Further, starting with one such solution, we can perturb the solution
by an arbitrary deformation that preserves the boundary values. We
have seen that the perturbed configuration relaxes back to the
original starting solution rapidly (in iteration time). This is an
indirect argument for the stability of the solution as well.

\subsection{Numerical Errors}

There are two main sources of errors in the relaxation method
described above.  First, there is the discretization error caused by
using a lattice description.  This may be countered in two obvious
ways.  We could use discrete representations of derivatives that are
of higher order in $h_x$ and $h_z$ and/or use a finer lattice. We
however resort to a version of the multigrid method. That is to say,
we first solve the numerical system on a coarse lattice and then use
this approximate solution to seed the starting configuration on a
finer lattice. This procedure converges much faster than naively
solving on the larger lattice. The second source of error is the
degree to which we solve the algebraic equations obtained after
discretizing the differential equations.  The most obvious way to
minimize the algebraic error is to simply let the system relax over
more iterations.

As a measure of the error in our numerical solutions, we define a
quantity which we call equation error, $\mathcal{E}$.  This is the
extent to which the numerically determined solutions do {\em not}
solve the equations of motion schematically written as
\beq
E.O.M(\tilde{R}_{num},A_{num}) = \mathcal{E}.
\eeq
To evaluate $\mathcal{E}$, we can again discretize the equations of
motion by now using n-th order representations for the various
derivatives and evaluate $\mathcal{E}_{num}(n)=E.O.M_n
(\tilde{R}_{num},A_{num})$. For a given numerical solution, we find
that $\mathcal{E}_{num}(n)$ hardly changes for $n>5$. We may therefore
use this numerically determined equation error (at large derivative
order$n>5$ ) as an approximation of the exact error in the numerical
solution
\beq
\mathcal{E}\sim\mathcal{E}_{num}(n)\Big{|}_{\mathrm{large}\ n}.
\eeq
In order to reduce the discretization error we may pass numerically
obtained solutions obtained on coarse lattices on to finer lattices
and repeat the process.  In this manner it is possible to
systematically reduce the equation error, $\mathcal{E}$.  In principle
this process could be continued indefinitely, although in reality we
are limited to lattices of size less than $513\times 513$ sites when
using Mathematica on a desktop computer.
\begin{figure}[ht]
\begin{center}
\includegraphics[totalheight=4cm]{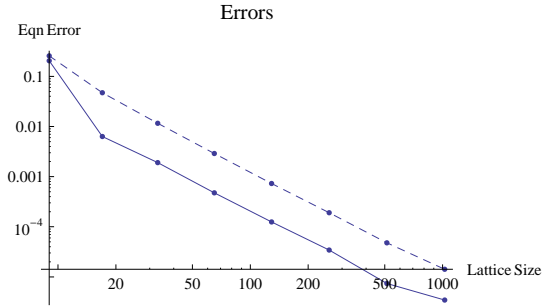}
\caption{$\mathcal{E}_{{\rm dis}}(7)$ as a function of lattice size.
  The dashed line is the maximum value of the $\tilde{R}$ equation of
  motion evaluated on our numerical
  solution and the solid line is the corresponding value for the
  gauge field}\label{fig:errors}
\end{center}
\end{figure}

\subsubsection{Uncertainty Analysis}

Although we have introduced a measure of numerical error in the
solutions, it would be more insightful to have a measure of the error
in physical quantities.  One way to do this comes from
(\ref{eq:variation}) which establishes the relationship between
normalizable modes and operator expectation values. Specifically, the
error in the condensate and charge density may be estimated by noting
that since the equations of motion do not actually evaluate to zero,
\beqn\label{eq:uncert}
 \Delta \left(\langle O_2\rangle \right)&\sim& \frac{1}{z_T^2}
\int d^4x {\Big |}\frac{\delta\tilde{R}}{\delta \tilde{R}^{(1)}}
{\Big |} |\mathcal{E}_{num}( \tilde{R})|
+\frac{1}{2}{\Big |}\frac{\delta A}{\delta \tilde{R}^{(1)}}
{\Big |}|\mathcal{E}_{num}( A)|
\\
\Delta \left(\langle \rho \rangle \right)&\sim& \int
\frac{d^4x}{f(z)}{\Big |}\frac{\delta\tilde{R}}{\delta A^{(0)}}
{\Big |} |\mathcal{E}_{num}( \tilde{R})|
+\frac{1}{2}{\Big |}\frac{\delta A}{\delta A^{(0)}}
{\Big |}|\mathcal{E}_{num}( A)|\eeqn
for the $\mathcal{O}_2$ condensate. Similar expressions may be
obtained for the $\mathcal{O}_1$ condensate.

As a final input we need to evaluate the functional derivative of bulk
fields with respect to their boundary values.  We will approximate the
functional derivatives by their values for the homogeneous solutions
(as determined numerically using Mathematica's NDSolve routine). That
is to say, we determine the change $\delta\tilde{R}(z)$ when we change
the boundary values by $\delta\tilde{R}^{(1)}$. We have checked that
this latter approximation is a good estimate of the actual functional
derivative $\frac{\delta\tilde{R}}{\delta \tilde{R}^{(1)}}$ even if we
use the full inhomogeneous equations.

Such as they are, these error estimates do not represent upper bounds
on the total error on spatially inhomogeneous solutions, but we do
believe that they are a descriptive of the error in our numerical
solutions.

\section{Dark Solitons\label{Dark Solitons}}

In order to identify quantities of interest to be extracted from our
numerical solutions, we will attempt to compare our holographic
solitons with the soliton solutions of the Gross-Pitaevskii(GP)
equation (the latter gives a coarse-grained description of a
superfluid valid at long wavelengths). However, we emphasize that we
will only be using the GP equation as a guide - the results indicate
that the holographic solitons are not the same as the solution of the
GP equation.  We also note that the description of a superfluid system
in terms of the GP equation alone is inadequate to capture the effects
of significant density depletion in the superfluid (perhaps due to
strong interaction effects) and also the effects of having a
significant non-condensate fraction (at say high temperatures). In our
results, we find that both these contributions seem to be present and
significant in the holographic superfluids.

The order parameter of superfluidity is a complex scalar field (also
called the condensate wave function)- for which one can write down
equation of motion called the Gross-Pitaevskii equation
\be
-\frac{1}{2m_B}\del^2\psi+(V-\m)\psi+ g\psi |\psi|^2=0 \label{GPeqns}
\ee
(we have dropped the time dependence in the above since we are only
interested in static phenomena in this work).

The dark soliton, in this language is a spatially varying solution of the
GP equation which interpolates between the potential minima
\be
\psi(x\to \infty)=\Delta \qquad \psi(x\to -\infty)= -\Delta
\ee
If we further assume translational invariance in the y-direction,
the GP equations (\ref{GPeqns}) has a well known exact solution
\beqn
\psi=\sqrt{\frac{V-\mu}{g}}\tanh(x/\xi),
\label{tanh}
\eeqn
which interpolates between the two minima at
$\psi=\pm\Delta=\sqrt{\frac{V-\mu}{g}}$.
The correlation length (or the healing length) $\xi$, can then be
written in a useful form in terms of the parameters of the
GP equation as
\beqn
\xi^2=\frac{1}{2g\Delta^2}=\frac{1}{(V-\mu)m_B}.\label{eq:GL5}
\eeqn
Although the GP equation is really relevant for a nonrelativistic
system, it will be interesting to test these dependences of the
coherence on the magnitude of the condensate and on the chemical
potential. Finally, because the charge density is simply related to
the order parameter $\rho\sim |\psi|^2$, the solution (\ref{tanh}) has
vanishing charge density at its core.

\subsection{Holographic results}

\begin{figure}[ht]
\begin{center}
\includegraphics[totalheight=3cm]{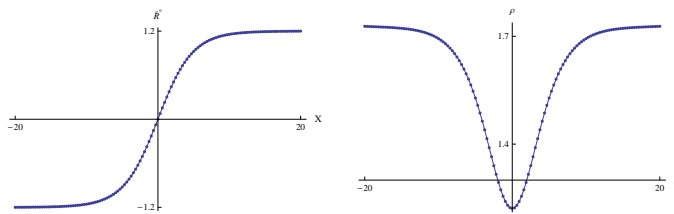}
\caption{$\oo_1$ Condensate and charge density as a function of $x$
}
\label{fig:O1fits}
\end{center}
\end{figure}
\begin{figure}[th]
\begin{center}
\includegraphics[totalheight=3cm]{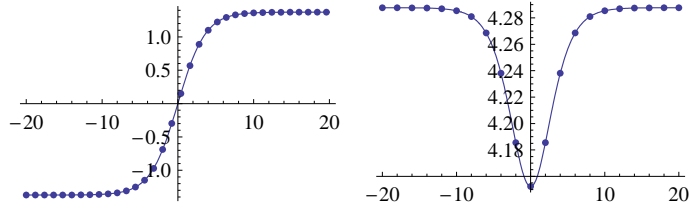}
\caption{$\oo_2$ Condensate and charge density as a function of $x$. }
\label{fig:O2fits}
\end{center}
\end{figure}

From the numerical solutions (\ref{fig:condo1},\ref{fig:condo2})
obtained by solving the {\em bulk equations} - we can extract the
boundary profiles of the charge density $A^{(1)}$ and the condensate
$R^{(1,2)}$ respectively. The plots as well as expectations from the
GP equation suggest that the condensate can be fitted by a
$\tanh(\frac{x}{\xi})$ profile and hence plausibly, the charge density
can be fitted by a $\sech^2(\frac{x}{\xi_q})$ profile. These data points
along with the best fit curve are shown in Fig. \ref{fig:O1fits} for
a typical $\oo_1$ type of condensate.  The results of a similar analysis for
the $\mathcal{O}_2$ system is shown in Fig. \ref{fig:O2fits}.  The
profiles of the boundary observables are equally well fit for both
$\oo_1$ and $\oo_2$ cases.
Using a least square fit, we then extract the coherence length $\xi$
from the condensate profile and $\xi_q$ from the charge density. It
must be noted here that the fit value of $\xi$ and $\xi_q$ are quite
sensitive to the accuracy of the solutions. We have ensured that the
equations are solved to an accuracy of $10^{-4}$ - that is to say, the
maximum absolute value of the right hand side of the differential
equation is less that $10^{-4}$.

\begin{figure}[th]
\begin{center}
\includegraphics[totalheight=5cm]{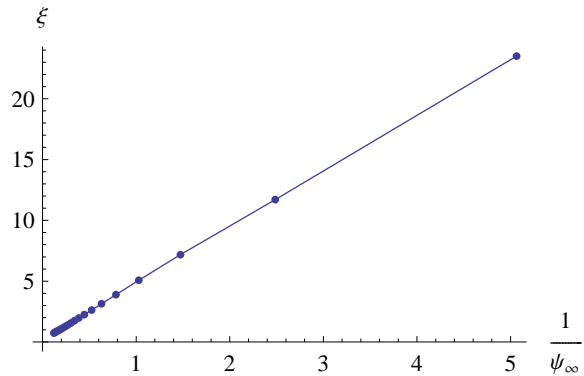}
\caption{$\oo_2$ Coherence length versus inverse condensate. }
\label{fig:xidelta}
\end{center}
\end{figure}

Following our earlier discussion of the GP-equation (\ref{eq:GL5}), we
plot the behaviour of the coherence length as a function of the
condensate in Fig. \ref{fig:xidelta}. One of the first surprises is
that $\xi$ is a linear function of the inverse condensate over a large
range of chemical potential, although there is no obvious reason for it in this context.

\subsection{Two Length Scales\label{2Length}}

\begin{figure}[ht]
\begin{center}
\includegraphics[totalheight=5cm]{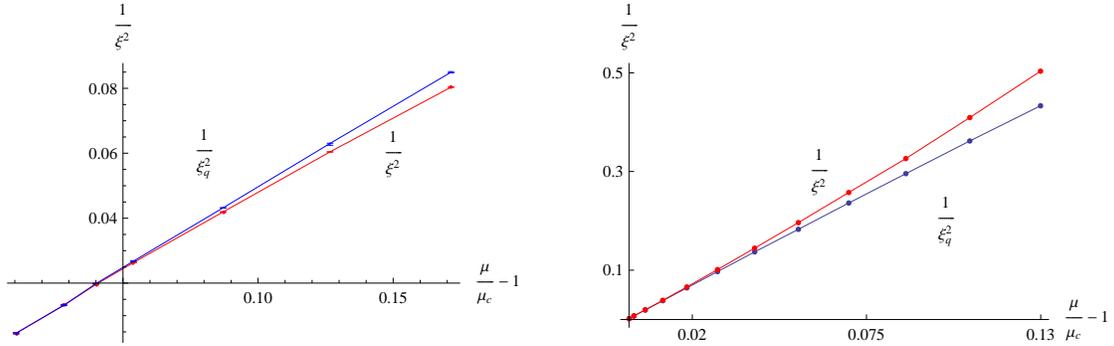}
\caption{The coherence lengths as a function of the chemical
  potential ($\oo_1$ on the left and $\oo_2$ on the right)}\label{fig:Figximu}
\end{center}
\end{figure}

Secondly, again motivated by (\ref{eq:GL5}), we would like to
determine the dependence of the coherence lengths $\xi$ on the
chemical potential $\mu$. This is shown in Fig. \ref{fig:Figximu}.
It is seen that for values of the chemical potential close to the
critical value (i.e., T close to $T_c$), these mass scales are linear
functions of the chemical potential (in this manuscript, we use the
term mass scale interchangeably with inverse length scale). This is
true for both the $\oo_1$ and $\oo_2$ solutions - but $\xi$ and $\xi_q$
have slightly different slopes as functions of the chemical potential.

\begin{figure}[ht]
\begin{center}
\includegraphics[totalheight=3cm]{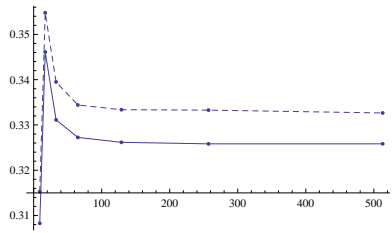}
\caption{ The best fit valued for $\xi$ (bottom) and $\xi_q$ (top)
  vs. lattice size for an $\oo_2$ condensate at
  $\frac{\mu}{\mu_c}=1.03$.}
\label{fig:xi-lat}
\end{center}
\end{figure}

This is in sharp contrast to what one might have expected from either
GP-equation or even the bulk equations. The numerical solutions
display different length scales for the condensate and charge density
(for $\mu>\mu_c$).
We have confirmed that this is not a numerical artefact in the
following ways. As a first check, we plot the behavior of the two
length scales as we increase lattice size in Fig. \ref{fig:xi-lat}.
We see that the difference between the $\xi$'s saturates for large
lattice sizes, indicating that errors due to discretization will not
swamp the difference between the two scales.

\begin{figure}[ht]
\begin{center}
\includegraphics[totalheight=5cm]{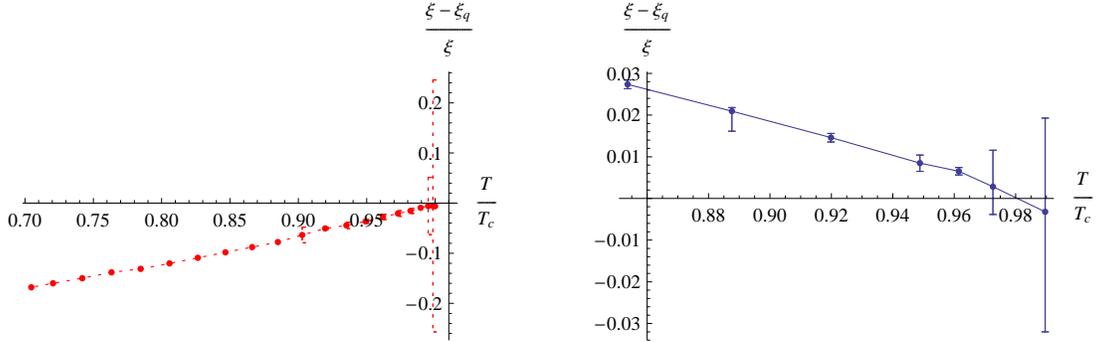}
\caption{$\frac{\xi-\xi_q}{\xi}$ vs. $\frac{T}{T_c}$: On the left is
  $\oo_2$ and the right is $\oo_1$}
\label{fig:diffxi}
\end{center}
\end{figure}

We also plot the relative difference between $\xi$ and $\xi_q$ as a
function of the chemical potential in Fig. \ref{fig:diffxi}.
Following the earlier discussion on error estimate for the condensate
and charge density profiles, we have determined error bars on these
numbers (which have also been indicated in Fig. \ref{fig:diffxi}).
Since the error bars do not overlap with zero (except close to $T_c$),
it is clear that the difference in these length scales is not a
numerical artefact.
A sharp difference between the two kinds of condensates is seen in the
observation that the length scale difference changes sign. 

The existence of two distinct length scales leads one to expect an
interesting layering effect as one nears the soliton's core.  Two
length scales would indicate that the relative fraction of charge
density in the condensate versus non-condensate degrees of freedom
varies as a function of $x$.  For $\oo_1$ we find that the condensate's
length scale is larger than that of the total charge density.  As one
approaches the soliton from infinity, the $\oo_1$ condensate starts to
drops off before the total charge density, therefore the
non-condensate contribution to the total charge density must be must
be relatively over dense when compared to its asymptotic value.  For
$\oo_2$, the relative order of the length scales is reversed, implying
that the density in the non-condensate fraction must be relatively
under dense near the dark soliton.

\subsection{Depletion fraction}
\begin{figure}[ht]
\begin{center}
\includegraphics[totalheight=5cm]{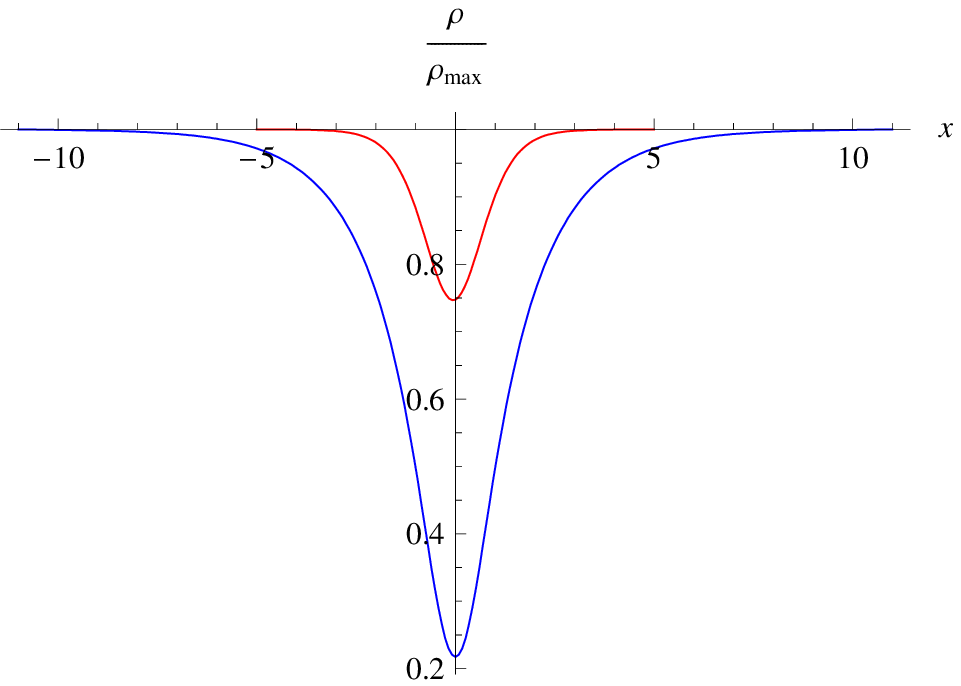}
\includegraphics[totalheight=5cm]{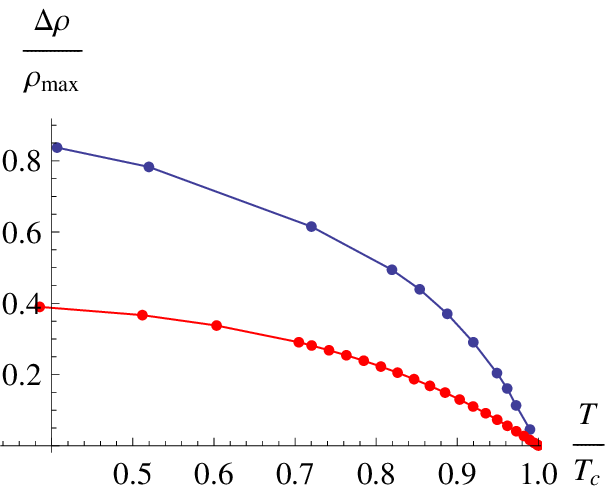}
\caption{Left: density depletion: $\oo_1$ is blue and dotted. Right:
  density depletion as a function of the temperature: $\oo_1$ on the
  top and $\oo_2$ on the bottom}
\label{fig:depfrac}
\end{center}
\end{figure}

Another quantity of interest in these objects is the amount of density
depletion at the core of the soliton. This is illustrated in the left
panel of Fig. \ref{fig:depfrac} where we plot the fractional density
as a function of the distance from the core of the soliton (for
$\frac{\mu}{\mu_c}=1.9$). One sees that there is a striking difference
in the amount of density depletion at the core between $\oo_1$ and
$\oo_2$ type of condensates.
Fig. \ref{fig:depfrac}.
In the right panel, we plot the percentage density depletion at the
core as a function of the chemical potential for the two kinds of
condensates. Firstly, the fact that this depletion fraction is not
100\% is an indicator that we are quite far away from the solution
(\ref{tanh}) of the GP equation.

In \cite{antezzaetal}, it was shown that the density depletion
fraction, at zero temperature, was directly related to whether the
system was BEC-like (large depletion) or BCS-like (small depletion)
(for non-relativistic systems). It is interesting to note that the
density depletion for holographic solitons in Fig. \ref{fig:depfrac}
is strongly dependent on the type of operator which condenses.  The
amount of density depletion for the $\oo_1$ condensate is likely to be
quite large near zero temperature. By contrast, $\oo_2$ type condensate
seems to be saturating near 40\% density depletion as we tend to zero
temperature.

If one defines the type of superfluid (BEC or BCS) by the amount of
density depletion at the core \cite{antezzaetal}, then
Fig. \ref{fig:depfrac} suggests that the $\oo_1$ system is likely to be
BEC type. However, as we shall see later, it is to be emphasized that
the comparison with the BCS/BEC types of superfluids is perhaps an
analogy only.

\subsection{Comparison of solitons with quasiparticles}
The dark soliton we have constructed is a ``macroscopic'' object in
this system with a characteristic length scale, namely the coherence
length. On the other hand, we also have quasiparticles which are
massive. These are ``microscopic'' excitations, and one could try to
compare the inverse mass of the quasi particle with the coherence
length. In a sense, since linear response is controlled by the
quasiparticle mass, it will be interesting to wonder if the coherence
length may be accounted for using some heuristic based on linear
response theory.

The quasiparticle mass can be determined by studying the two point
functions for fluctuations around the homogeneous condensate, and
examining the fall-off as a function of the {\em spatial} co-ordinate
x (we {\em define} the quasi-particle mass this way).  The boundary
conditions satisfied by the fluctuations in this case are
``reflective'' regularity conditions at the horizon.  This is
different from the sound mode studies in that the latter uses
infalling boundary at the horizon as is appropriate for a black hole
quasi-normal mode (which represents a relaxation mode for the
superfluid system).

Using the AdS/CFT dictionary, the boundary quasiparticle masses arise
from the poles in the bulk to boundary propagator in Euclidean AdS
space\cite{Witten:1998zw}.  Because we want the lightest masses
(largest length scales), it is only necessary to solve for the lowest
pole which has vanishing Matsubara frequency. Finding the lowest pole
of the bulk to boundary propagator reduces to finding the values of
$k^2$ for which one has a static solution to the linearized equation
of motion satisfying the correct boundary conditions (see below).

In detail, we can linearize (\ref{eq:bulk}) to find
\bea
f \delta R''+f' \delta R'-z \delta R -k^2\delta R +\delta R
(\frac{A^2}{f})+\frac{2A\tilde R}{f} \delta A=0\\
f \delta A''-k^2\delta A-\tilde R ^2 \delta A-2A\tilde R \delta A=0
\label{eq:fluct}
\eea 
where $k$ represents the (dimensionless) momentum of the particle in
the x-direction and we have dropped the tilde on the fluctuation
$\delta \tilde R(k,z)$ ($\tilde R,A$ are solutions representing the
background). The presence of the interation term between $\delta
\tilde R$ and $\delta A$ suggests that there is mixing between these
two kinds of quasiparticles - very similar to the mixing between axial
and vector-mesons in models of AdS/QCD \cite{Domokos:2007kt}.

The boundary conditions at the horizon are again obtained
by requiring regularity of the solutions
\be
-3\delta R'(k,z=1)=(1+k^2)\,\delta R(k,z=1) \qquad \delta A(k,z=1)=0
\ee
One may understand this as a requirement that the excitation energy
of the quasiparticle over the ground state energy be finite.

\begin{figure}
\begin{center}
\includegraphics[totalheight=5cm]{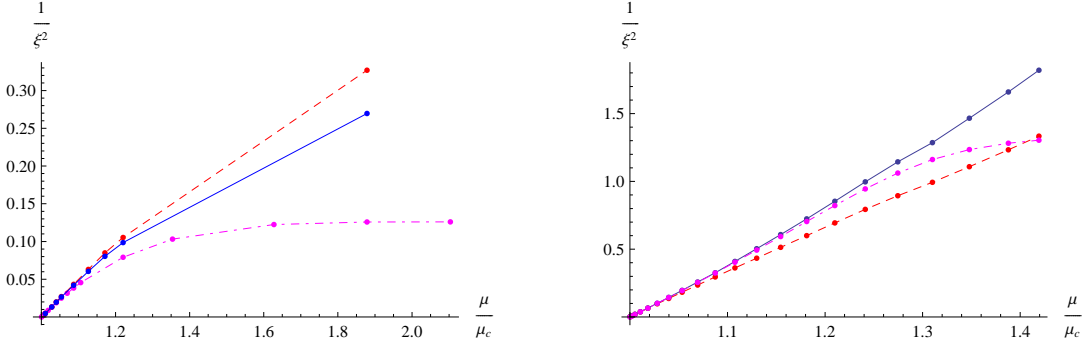}
\caption{Quasiparticle mass (magenta, dot-dashed) as a function of
  $\mu$ with $\xi$ (blue) and $\xi_q$ (red, dashed)}: $\oo_1$ on the left and $\oo_2$ on the right.
\label{qp-mass}
\end{center}
\end{figure}

In order to determine the boundary conditions to be imposed at the AdS
boundary, we again determine the behavior of the modes close to the
boundary and require that the corresponding ``non-normalizable'' modes
are zero. This gives
\be
\delta R(z=0)=0\qquad \delta A(z=0)=0
\ee
We now solve the equations (\ref{eq:fluct}) subject to the above
boundary condition as an {\em eigenvalue} problem for $M^2=-k^2$.  For
this value of $k^2$ there will be a vanishing eigenvalue for the
linearized equations of motion, and hence a pole in the bulk to
boundary propagator.  The resultant graph of quasiparticle mass
vs. temperature is shown in Fig. \ref{qp-mass} along with the
coherence lengths.
It is interesting to note that for $\oo_2$ the quasiparticle length
scale tracks the soliton's condensate length scale for
$\frac{\mu}{\mu_c}<1.2$.  This is surprising in that one might have
expected the lightest quasiparticle to follow the largest length
scale.  This might be a sign that one should also include other fields
in the quasiparticle analysis (in the sense that one of the other
fields has poles which track the larger length scale). To date, our
numerical accuracy is insufficient to see if this is also true for
$\oo_1$.

\section{Discussion}
In this work, we have studied properties of one type of extended
configuration allowed in the holographic superfluids described in
\cite{Hartnoll:2008vx}.  These superfluids support dark soliton
solutions which are characterized by a local depletion in the charge
density.  This paper explored various features of these solitons as
functions of the chemical potential and condensate type using
numerical simulations.

Although the system we actually solved is a Maxwell-Higgs system in
the presence of a black hole, the magic of the holographic AdS/CFT
correspondence results in a dark soliton with parallels to the soliton
of the GP equation.
In fact, we found that the length scale associated with the soliton
scales with the chemical potential as for soliton of the GP equation.

However, we found that the variation in the order parameter occurs
with a different length scale than for the charge density.  This is a
feature for superfluids of both $\bra \oo_1\ket$ and $\bra \oo_2\ket$
types.  The presence of these two length scales is a surprising result
which is not manifest in the original gravitational system nor was it
predicted by hydrodynamical studies.

A feature of these scales is that for the $\bra \oo_1\ket$ superfluid,
the length scale associated with the order parameter is larger than
that associated with the total charge density.  As discussed in
Section \ref{2Length}, this indicates an interesting spatial
dependence of the non-condensate fraction of the charge density. If
we use the size of order parameter as an indicator of the fraction of
charge density which resides in the condensate itself, we can conclude
that the non-condensate fraction of the total charge density is over
dense near the dark soliton.  For the $\bra \oo_2\ket $ superfluid, the
relative size of the two length scales is reversed and we find that
the non-condensate fraction must be under dense near the soliton core.

Spatial ordering in the components of the charge density will almost
certainly affect transport phenomena near the dark soliton.  The
presence of different spatial orderings in the non-condensate charge
density might allow one to test how the quasiparticles see the
condensate differently than the non-condensate matter, although it
should be pointed out that the relative difference in the two length
scales is small for the chemical potentials discussed in this paper.
This would be an interesting direction to pursue in future work.

A second feature that we found is that, at finite temperature, the
density depletion in the core of the soliton is typically quite far
from 100\%. In this sense too, the solitons of the GP equations are a bad
guide.  The actual value of the density depletion fraction
strongly depends upon the type of holographic superfluid studied.  For
an $\bra \oo_1\ket$ condensate we find that the depletion fraction grows to
near 100\% for low temperatures, whereas for an $\bra \oo_2\ket$
condensate the depletion fraction seems to asymptote towards 40\% for
low temperatures.

These same gross features are known to exist in non-relativistic
superfluids.  In the context of the BEC-BCS crossover, it is known
that BEC dark solitons have near 100\% depletion fractions, while BCS
superfluids have depletion fractions less than 60\%
\cite{antezzaetal}.  In that case, the difference is associated with
the characteristic size of the field which condenses.  In the BEC
regime it is a pointlike boson condensing while in the BCS regime the
condensate is comprised of large Cooper pairs.

It should be noted that in holographic superfluids, the precise
microscopic description of the system is not known. In general, it may
be expected that we can construct both fermionic and bosonic
quasiparticles.  One could ask if an analogous classification of the
size of the quasiparticles (into tightly bound BEC atoms and
relatively large sized Cooper pairs) is possible in {\em relativistic}
holographic superfluids.  Therefore, we could hope to realize features
of both types of both BEC and BCS type superfluids by tuning
parameters of the gravitational description.

If we associate the depletion fraction of dark soliton with the
scaling dimension of the condensing operator, it is natural to imagine
that one may change the depletion fraction as one tunes the scalar
mass in the gravitational action.  In fact, in the works of
\cite{Umeh:2009ea, Kim:2009kb}, the authors found a rich structure of
ground states as the mass of the bulk scalar varied away from
$L^2m^2=-2$.  It would be quite interesting to explore how the
properties of the quasiparticles and solitons vary with $m^2$ and
whether one can find a crossover analogous to what is found for
non-relativisitc superfluids.


Of course, one of the important question to be answered before taking
the analogy to the BCS system seriously is whether there is a Fermi
surface, or some aspect of a Fermi surface in the dual system.  There
have been several explorations in this regard and in fact, it has even
been studied if one can realize a (non-)Fermi {\em liquid} holographically
\cite{Lee:2008xf,Liu:2009dm,Faulkner:2009wj,Cubrovic:2009ye,
  Albash:2009wz,Basu:2009qz}.

One potential shortcoming of our work is the fact that we do not have
a smooth limit to zero temperature. This is an artefact of the probe
limit\cite{Horowitz:2009ij}. Therefore, a numerically challenging
problem which is potentially of much interest is to construct the
fully backreacted black hole with a dark soliton. Such a solution to
the gravity system would be very interesting in the sense that it
would give this black hole rather novel hair.  For a recent discussion
along this line see \cite{Nakamura:2009tf} (of course, in string
theory contexts, such spatially inhomogeneous hair have a complex
history).

We have observed (\ref{recap}) that the presence of the condensate
perturbs the gauge field background by little. Therefore, it might be
interesting to explore a different pertubation expansion, by
considering scalar fields in a charged black hole background. This
would amount to including the backreaction of the noncondensate part,
and thus correspond to an expansion in the superfluid density.

We also note that in recent
works\cite{Albash:2009iq,Montull:2009fe,Maeda:2009vf}, several other
extended solutions of the gravitational system were studied. In these
works, non-normalizable components of some bulk fields were turned on
and the solitons were then interpreted as being vortices in a
superconductor. In this sense, while these authors study the same set
of equations, their results are interpreted quite differently. In a
subsequent work, we will present numerical results about vortices in a
supefluid. The main difference with the previous work is the absence
of any non-normalizable component to the bulk magnetic field.

Finally, one would like to explore such phenomena in the context of
non-relativistic gauge gravity duals. In this way one might hope to
model the experimentally observed dark solitons much more closely.

\section{Acknowledgements}
We would like to thank Lincoln Carr, Ari Harju and Smitha Vishveshwara
for useful discussions.  The numerical results presented in this work
were obtained by extensive programming in Mathematica. We wish to
acknowledge our debt to the makers of this software. V.K. and
E.K-V. have been supported in part by the Academy of Finland grant
number 1127482.

\end{document}